\documentclass[10pt]{article}
\usepackage[utf8]{inputenc}
\usepackage{graphicx}
\usepackage{dcolumn}
\usepackage{bm}
\usepackage{chngpage}
\usepackage{graphicx}
\usepackage{braket}
\usepackage[margin=0.9in]{geometry}
\usepackage{amsmath}
\usepackage{float}
\usepackage{graphicx}
\usepackage{caption}
\usepackage{amsmath, amssymb}
\usepackage{bbold}
\usepackage{makecell}
\usepackage{braket}
\usepackage{calc}
\usepackage{subcaption}
\usepackage{authblk}
\usepackage{indentfirst}
\usepackage{lipsum}
\usepackage{marvosym}
\usepackage{lineno}
\usepackage{soul}
\usepackage[version=4]{mhchem}

\usepackage[
autocite=superscript,
backend=biber,
style=nature,
date=year,
doi=false,isbn=false,url=false,eprint=false
]{biblatex}
\usepackage[lang=en]{jabbrv}
\AtEveryBibitem{\clearlist{language}}
\usepackage[labelfont=bf]{caption}
\usepackage{amsmath}
\usepackage[dvipsnames]{xcolor}
\usepackage{geometry}
 \usepackage[%
    colorlinks=true,
    pdfborder={0 0 0}{},
    linkcolor=blue,
    citecolor=blue,
]{hyperref}
\usepackage{textcomp, gensymb}
\DefineJournalPartialAbbreviation{Rhapsod}{Rhaps}

\title{\textbf{Spontaneous emergence of phonon angular momentum through hybridization with magnons}}

\author[1,*]{Honglie Ning}
\author[1,*]{Tianchuang Luo}
\author[1,*]{Batyr Ilyas}
\author[2,3]{Emil Viñas Boström}
\author[4]{Jaena Park}
\author[4]{Junghyun Kim}
\author[4]{Je-Geun Park}
\author[5,6]{Dominik M. Juraschek}
\author[2,7]{Angel Rubio}
\author[1,\Letter]{Nuh Gedik}

\affil[1]{Department of Physics, Massachusetts Institute of Technology, Cambridge, 02139, Massachusetts, USA.}
\affil[2]{Max Planck Institute for the Structure and Dynamics of Matter, Luruper Chaussee 149, 22761 Hamburg, Germany.}
\affil[3]{Nano-Bio Spectroscopy Group, Departamento de Fisica de Materiales, Universidad del Pais Vasco, 20018 San Sebastian, Spain.}
\affil[4]{Department of Physics and Astronomy and Institute of Applied Physics, Seoul National University, Seoul 08826, Republic of Korea.}
\affil[5]{School of Physics and Astronomy, Tel Aviv University, Tel Aviv 6997801, Israel.}
\affil[6]{Department of Applied Physics and Science Education, Eindhoven University of Technology, Eindhoven, Netherlands}
\affil[7]{Center for Computational Quantum Physics, The Flatiron Institute, New York, NY 10010, USA.}
\affil[*]{These authors contributed equally to this work.}
\affil[\Letter]{e-mail: gedik@mit.edu}

\begin{document}
\maketitle

\section*{Abstract}
Chirality, the breaking of improper rotational symmetry, is a fundamental concept spanning diverse scientific domains.
In condensed matter physics, chiral phonons, originating from circular atomic motions that carry angular momentum, have sparked intense interest due to their coupling to magnetic degrees of freedom, enabling potential phonon-controlled spintronics.
However, modes and their counter-rotating counterparts are typically degenerate at the Brillouin zone center. 
Selective excitation of a single-handed circulating phonon requires external stimuli that break the degeneracy.
Whether energetically nondegenerate circularly polarized phonons can appear spontaneously without structural or external symmetry breaking remains an open question.
Here, we demonstrate that nondegenerate elliptically polarized phonon pairs can be induced by coupling to  magnons with same helicity in the van der Waals antiferromagnet $\mathrm{FePSe_3}$.
We confirm the presence of magnon-phonon hybrids, also known as magnon polarons, which exhibit inherent elliptical polarization with opposite helicities and distinct energies.
This nondegeneracy enables their coherent excitation with linearly polarized terahertz pulses, which also endows these rotating modes with chirality.
By tuning the polarization of the terahertz drive and measuring phase-resolved polarimetry of the resulting coherent oscillations, we determine the ellipticity and map the trajectory of these hybrid quasiparticles.
Our findings establish a general approach to search for intrinsically nondegenerate phonons with angular momentum at the center of the Brillouin zone and introduce a new methodology for characterizing their ellipticity, outlining a roadmap towards chiral-phonon-controlled spintronic functionalities.
\section*{Main Text}

\paragraph{}
Asymmetric population of opposite chiralities is fundamentally prevalent across a vast array of systems, ranging from parity violation in weak nuclear interactions to the homochirality of biomolecules like DNA\autocite{Cahn1966SpecificationChirality,Barron1986SymmetryChirality,Hasan2017DiscoveryStates,Schaibley2016ValleytronicsMaterials}{}.
In condensed matter physics, chirality emerges in crystals that lack all improper rotational symmetries. 
These structures naturally host exotic excitations such as Weyl fermions and circularly polarized phonons, with the latter garnering extensive attention recently, as their circular atomic motion carries angular momentum that can couple to electronic degrees of freedom, producing a plethora of unconventional transport and optical phenomena\autocite{Zhang2015ChiralLattices,Zhu2018ObservationPhonons,Chen2019EntanglementWSe2,Geilhufe2023ElectronKTaO3,Zhang2022ChiralMaterials,Zhang2022ChiralSystems,Ishito2023Truly-HgS,Ueda2023ChiralX-rays,Ren2021PhononMagnetization,Hernandez2023ObservationTopology,Chaudhary2024GiantCoupling,Ren2024Light-DrivenParamagnets,Juraschek2019OrbitalPhonons,Juraschek2020Phono-magneticEffects,Juraschek2022GiantParamagnets,Kahana2023Light-inducedRectification,Dornes2019TheEffect,Tauchert2022PolarizedDemagnetization,Grissonnanche2019GiantSuperconductors,Grissonnanche2020ChiralCuprates,Chen2022LargeCu3TeO6}{}.
Intriguingly, these modes have been observed not only in inherently chiral materials but also in nominally achiral systems.
In such crystals, however, selective excitation of a single-handed mode typically requires external stimuli that break the degeneracy between the two modes with opposite handedness, such as circularly polarized light or magnetic fields aligned with the phonon angular momenta \autocite{Baydin2022MagneticPbTe,Cheng2020ASemimetal,Cui2023ChiralityAntiferromagnet,Lujan2024SpinorbitMagnet,Wu2023Fluctuation-enhancedAntiferromagnet,Disa2020PolarizingField,Luo2023LargeHalides,Davies2024PhononicEffect,Basini2024TerahertzSrTiO3}{}.
These external perturbations also determine the helicity of the excited phonon.
To date, whether phonon ellipticity can spontaneously emerge without structural or externally imposed symmetry breaking awaits experimental verification.
Discovering such systems would not only advance our fundamental understanding of these quasiparticles but also enable their coherent excitation using light of arbitrary polarization, potentially unlocking new chiral functionalities in materials science and technology.

\paragraph{}
One potential route to realize such nondegenerate elliptical phonons is to exploit the interaction between phonons and other quasiparticles possessing angular momenta. 
Magnons emerge as prominent candidates, as the ellipticity of their inherent spin procession can, in principle, be transferred to selectively coupled phonons while abiding by symmetry constraints \autocite{Hamada2020ConversionRotation,Wang2024MagneticMagnons}. 
Here, we demonstrate this scenario in the quasi-two-dimensional van der Waals antiferromagnet $\mathrm{FePSe_3}$, which hosts strong spin-lattice coupling, facilitating magnon-phonon interactions. 

\begin{center}
\begin{figure}[H]
   \sbox0{\includegraphics[width=\textwidth]{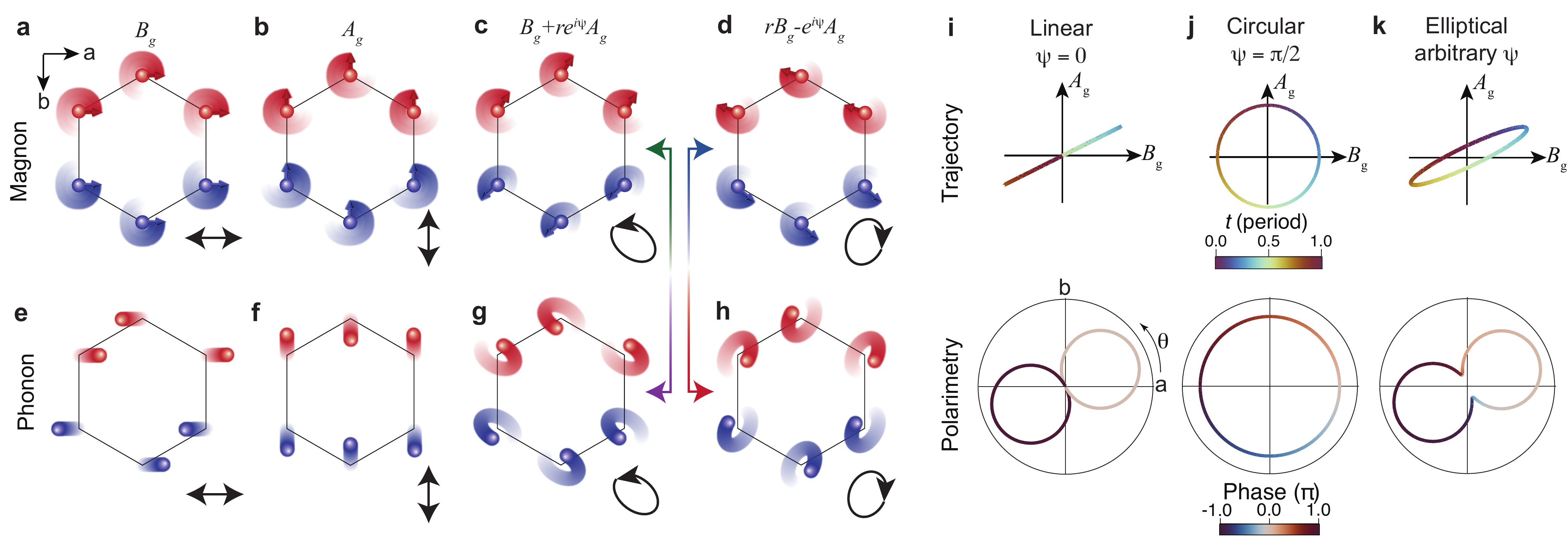}}
    \begin{minipage}{\wd0}
  \usebox0
  \captionsetup{labelfont={bf},labelformat={default},labelsep=period,name={Fig.}}
  \caption{\textbf{Trajectory and polarimetry of the elliptical magnon polarons.} \textbf{a-d}, Schematics of the trajectories of $B_g$ magnon, $A_g$ magnon, and the two orthogonal elliptical magnons composed of $B_g$ and $A_g$ magnons with a phase retardation. Net magnetization polarization direction is schematically shown at the corner of each panel. \textbf{e-h}, Schematics of the trajectories of $B_g$ phonon, $A_g$ phonon, and the two orthogonal elliptical phonons. Atomic motion direction is schematically shown at the corner of each panel. Selective coupling between the elliptical magnons and phonons are labeled by the colored double arrows. \textbf{i-k}, Top panels: schematics of the simplified magnetization/phonon trajectory of linear ($\psi=0$), circular ($r=1$,$\psi=\frac{\pi}{2}$), and elliptical ($\psi\neq0,\pm\frac{\pi}{2},\pi$) magnon polarons in $\{B_g,A_g\}$ coordinates with time $t$ color coded. Bottom panels: driving field polarization angle $\theta$-dependence of the amplitudes and phases of different modes. } 
  \label{Fig1}
\end{minipage}
\end{figure}  
\end{center}

\paragraph{}
$\mathrm{FePSe_3}$ comprises hexagonal Fe layers that adopt a zigzag antiferromagnetic pattern along the crystallographic $a$ axis with an easy axis along the out-of-plane direction below $T_\mathrm{N}\sim110$ K \autocite{Wiedenmann1981NeutronFePSe3,Xie2023IdentificationFePSe3,Ni2023SignaturesAntiferromagnets,LeMardele2024TransverseFePSe3,Chen2024ThermalFePSe3}{}. 
The leading-order magnetic Hamiltonian (Supplementary Notes 1 and 8) describes two degenerate magnons $m_B$ and $m_A$ around 3.5 THz.
These magnons respect $B_g$ and $A_g$ symmetries and feature linear magnetization oscillations along the crystallographic $a$ and $b$ axes, respectively (Figs.~\ref{Fig1}a,b)\autocite{Cui2023ChiralityAntiferromagnet,Luo2023EvidenceLimit}{}. 
Subdominant anisotropic exchange interactions or in-plane magnetic anisotropies can weakly break the degeneracy of these modes, yielding two new eigenmodes $m_1$ and $m_2$, each involving a mixture of $m_B$ and $m_A$\autocite{Jana2023In-planeScattering,Luo2023EvidenceLimit}{}. 
Without loss of generality, we express $m_1$ as $m_B+re^{i\psi}m_A$, assuming that $r$ is the ratio of the composition of $m_A$ to $m_B$ and $\psi$ is the phase retardation. 
To ensure orthogonality, $m_2$ must follow the form of $rm_B-e^{i\psi}m_A$. 
When $r\neq 0$ and $\psi\neq n\pi$ ($n\in \mathbb{Z}$), $m_1$ and $m_2$ represent two nondegenerate elliptically polarized magnons, with net magnetization circulating in orthogonal elliptical trajectories unaligned with any crystallographic axes and rotating in opposite directions (Figs.~\ref{Fig1}c,d). 
Meanwhile, a phonon with $B_g$ symmetry, $p_B$, and a phonon with $A_g$ symmetry, $p_A$, which emerge from the twofold degenerate $E_g$ mode below $T_\mathrm{N}$, are located in close vicinity to these magnons (Figs.~\ref{Fig1}e,f)\autocite{Cui2023ChiralityAntiferromagnet,Scagliotti1987RamanCrystals,Luo2023EvidenceLimit}{}. 
Previous experiments have demonstrated selective coupling between these nearly degenerate magnons and phonons, creating magnon-phonon hybrids known as magnon-polarons (MPs)\autocite{Cui2023ChiralityAntiferromagnet,Jana2023In-planeScattering,Luo2023EvidenceLimit,Zhang2021CoherentInsulator,Vaclavkova2021MagnonFePS3,McCreary2020Quasi-two-dimensionalSpectroscopy,Liu2021DirectFields}{}. 
Symmetry constraints dictate that such selective coupling can only happen between phonons and magnons composed of the same eigenvectors. 
This indicates that the new phonon eigenbases mirror the elliptically polarized magnons, i.e. $p_1=p_B+re^{i\psi} p_A$ and $p_2=rp_B-e^{i\psi}p_A$ (Figs.~\ref{Fig1}g,h) (Supplementary Notes 1). 
Consequently, $p_1$ and $m_1$ will selectively couple, forming MPs with left-handedness, while the MPs formed by $p_2$ and $m_2$ are right-handed, existing at distinct energies.
$\mathrm{FePSe_3}$ thus provides a unique platform where left- and right-handed MPs possess distinct energies even at the Brillouin zone center and under zero magnetic field, holding prospects for control using linearly polarized optical excitation.

\paragraph{}
A nondegenerate MP with a particular handedness in the terahertz (THz) spectral range can be resonantly launched by a linearly polarized THz pulse. 
The tiny but finite linear momentum transfer from the THz drive imparts chirality to these circulating MPs.
The magnetic component of the THz driving field can stimulate the magnonic part of the MP through magnetic dipolar interactions when its energy matches the MP energy\autocite{Kampfrath2011CoherentWaves,Mashkovich2021TerahertzLattice}{}.
Since linearly polarized light can be decomposed into left- and right-handed components and the chiral MP excitation is helicity selective, the component sharing the same handedness as the MP excites the hybrid mode, while the other component transmits through the sample, conserving the net angular momentum.
Consequently, by employing an intense linearly polarized broad-band THz pulse, multiple MPs with opposite chiralities that fall within the pulse bandwidth can be simultaneously excited.

\paragraph{}
The ellipticity of the these hybrid modes can be determined by measuring their amplitudes and phases as a function of the polarization angle $\theta$ of the linearly polarized THz pulse with respect to the crystal orientation. 
In the magnetic dipolar excitation channel, the MP amplitude is maximized when the THz driving magnetic field aligns with the magnetization generated by the magnon\autocite{Ilyas2023TerahertzFePS3,Luo2024ControllingAntiferromagnet,Zhang2024TerahertzAntiferromagnet,Zhang2024Terahertz-field-drivenAntiferromagnet}{}. 
Therefore, for a linearly polarized MP composed of $m_B$ and $m_A$ with $\psi=0$, the $\theta$-dependence of its amplitude forms a polar pattern with two petals, whose orientation, depending on $r$, deviates from the crytallographic axes (Fig.~\ref{Fig1}i).
Moreover, the coherent oscillation initiates in opposite directions as the driving field is rotated by $\pi$, causing the two petals to exhibit opposite phases that switch instantaneously from $0$ to $\pi$ at the node (Fig.~\ref{Fig1}i). 
On the other hand, a circularly polarized MP featuring $r=1$, $\psi=\frac{\pi}{2}$ creates a uniform circulating magnetic moment, resulting in a nodeless polar pattern with continuous phase change as a function of $\theta$ (Fig.~\ref{Fig1}j). 
In a more general context of elliptically polarized MP with arbitrary $r$ and $\psi$, a polar pattern incorporating features of both linearly and circularly polarized MPs should be otherwise expected: a nodeless ``peanut"-like polar pattern with the ``waist" rotated away from the crystallographic axes (Fig.~\ref{Fig1}k). 
Furthermore, the phase remains stable across most of the petal, but changes relatively abruptly around the waist. 
Given these distinct polar patterns of MPs with different ellipticities, phase-resolved polarimetry measurements can deterministically unveil the values of $\psi$ and $r$ for them.

\begin{center}
\begin{figure}[H]
   \sbox0{\includegraphics[width=\textwidth]{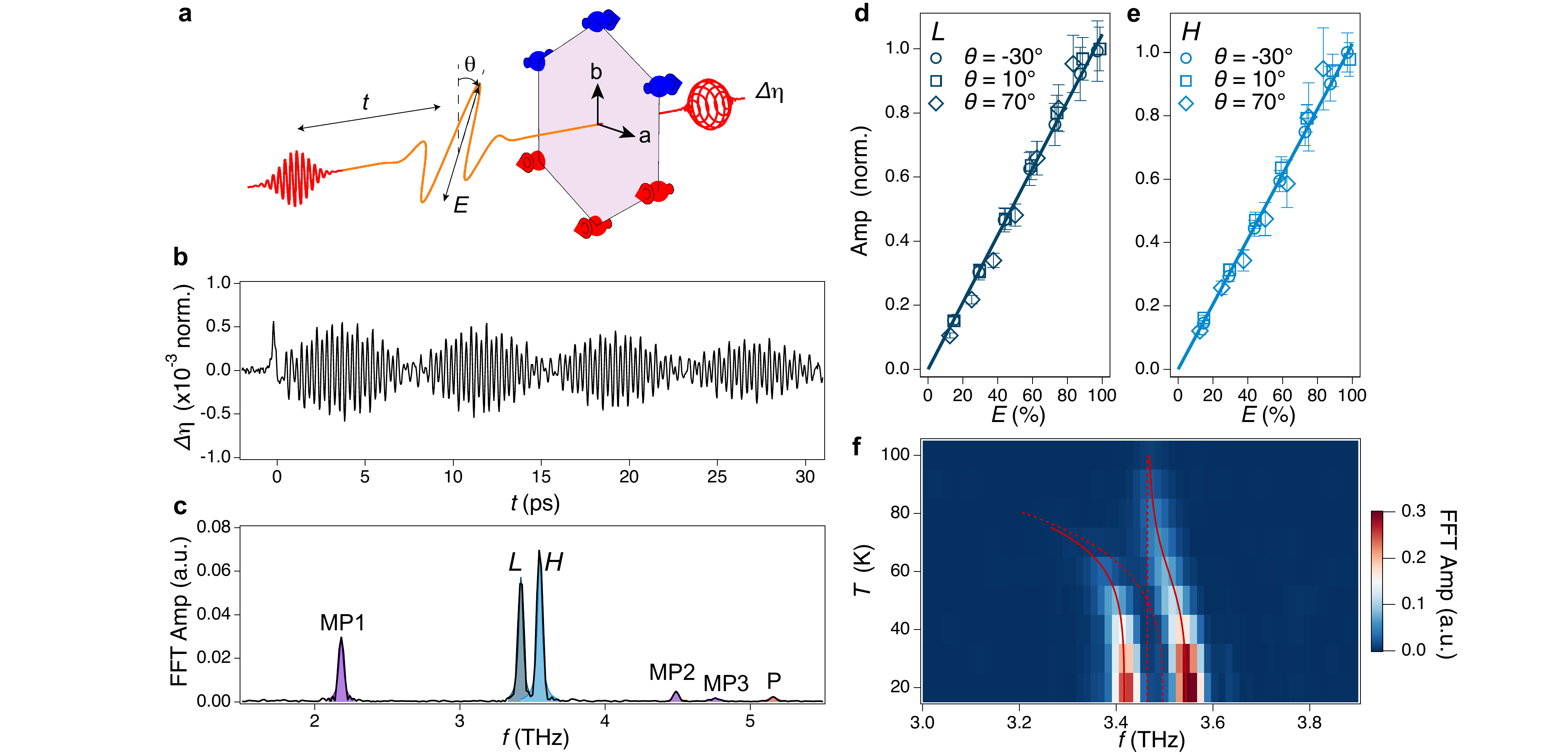}}
    \begin{minipage}{\wd0}
  \usebox0
  \captionsetup{labelfont={bf},labelformat={default},labelsep=period,name={Fig.}}
  \caption{\textbf{Driving field strength and temperature dependencies of the elliptical magnon polarons.} \textbf{a}, Schematic of the experimental setup. An intense THz pump (yellow) and a weak near-IR probe (red) are focused on the (001) surface of $\mathrm{FePSe_3}$ with a tunable time delay $t$. Both the driving field strength $E$ and polarization angle $\theta$ can be tuned (Methods and Supplementary Note 2). Polarization ellipticity change ($\Delta\eta$) of the transmitted probe is detected. \textbf{b}, $\Delta\eta$ transient obtained at 10 K at maximal $E$ ($\sim$150 kV/cm) and $\theta=150^\circ$. \textbf{c}, FFT spectrum of the time trace in \textbf{b}. The shaded regions are fits to each peak with a damped oscillator. $L$ and $H$ are both elliptical MPs, colored by dark and light blue, respectively. The other MPs (MP1, MP2, and MP3) and the uncoupled Raman-active phonon (P), colored by purple and red, respectively, are discussed in Supplementary Notes 4-6. \textbf{d,e}, Amplitudes of $L$ and $H$ MPs as a function of $E$ at different $\theta$ values. Data are normalized for a better comparison and fit with a linear function. \textbf{f}, Zoom-in FFT spectra around $L$ and $H$ MPs as a function of $T$ at maximal $E$ and $\theta=-30^\circ$. Solid lines denote the frequencies of the coupled MPs while dashed lines denote the frequencies of uncoupled bare magnon and phonon.} 
  \label{Fig2}
\end{minipage}
\end{figure}  
\end{center}

\paragraph{}
To excite the targeted elliptical MPs in $\mathrm{FePSe_3}$, we employ a high-field, ultrashort linearly polarized THz pulse covering a broad bandwidth from 0.5 THz to 6 THz (Supplementary Notes 2 and Methods), and control its peak electric field strength $E$ (proportional to its magnetic field $B$) and polarization angle $\theta$ independently (Fig.~\ref{Fig2}a). 
The THz-induced polarization ellipticity change ($\Delta\eta$) of a transmitted 800 nm probe pulse is tracked as a function of the relative time delay $t$, with MPs manifest as a series of coherent oscillations. 
A typical $\Delta\eta$ transient acquired at a temperature $T=10$ K, far below $T_\mathrm{N}$, is depicted in Fig.~\ref{Fig2}b. 
Ensuing the initial positive signal around $t=0$ stemming from the THz Kerr effect, beating patterns emerge, suggesting the coexistence of multiple frequency components. 
A fast Fourier transform (FFT) of the time trace in Fig.~\ref{Fig2}b reveals at least six Raman-active modes (Fig.~\ref{Fig2}c). 
The modes labelled as $L$ and $H$ at 3.4 THz and 3.55 THz, each of which is twofold degenerate, are the targeted elliptical MPs ($m_1$,$m_2$,$p_1$,$p_2$), as demonstrated in detail as follows.
Comprehensive characterization of the other modes is discussed in Supplementary Notes 4-6.

\paragraph{}
To reveal the magnonic nature of $L$ and $H$, we measure their amplitudes as a function of the peak electric field $E$ of the THz pulse. 
Since MPs host both magnon and phonon components, both magnetic dipolar and Raman active excitation channels should be activated, which are linear and quadratic in $E$, respectively (Supplemental Note 3)\autocite{Ilyas2023TerahertzFePS3,Luo2024ControllingAntiferromagnet}. 
In contrast to the other modes, which exhibit a combination of linear and quadratic $E$-dependencies (Supplementary Note 4), the amplitudes of $L$ and $H$ consistently display a linear dependence on $E$ at various $\theta$, indicating the predominance of the magnetic dipolar excitation channel (Figs.~\ref{Fig2}d,e). 
This highlights the magnonic nature of $L$ and $H$, in agreement with previous reports\autocite{Cui2023ChiralityAntiferromagnet,Jana2023In-planeScattering,Luo2023EvidenceLimit}. 

\paragraph{}
Since the nonlinear excitation channel is subdominant, whether $L$ and $H$ comprise phonons cannot be revealed by a measurement of the $E$-dependence. 
Temperature dependence provides an alternative pathway, as magnons soften significantly as $T$ increases towards $T_\mathrm{N}$, while phonon energies remain mostly unchanged\autocite{Cui2023ChiralityAntiferromagnet,Jana2023In-planeScattering}{}. 
Therefore, if $L$ and $H$ are MPs, we expect the degeneracy of their magnonic and phononic components to be progressively lifted as temperature increases.
To verify this hypothesis, we increase $T$ from 10 K to 100 K, slightly below $T_\mathrm{N}$, and examine the FFT spectrum evolution around $L$ and $H$ (Fig.~\ref{Fig2}f). 
We find that $L$ quickly loses its spectral weight above 50 K, accompanied by a dramatic frequency decrease of more than 0.15 THz, until it becomes unresolvable above 80 K. 
At the same time, despite a rapid loss of spectral weight, $H$ remains detectable up to 100 K, with its frequency saturating to $\sim3.48$ THz above 70 K. 
On the contrary, other modes only exhibit mild $0.01\sim0.02$ THz softening across the entire temperature range (Supplementary Note 6). 
We employ a toy model assuming coupling between a magnon whose energy exhibits an order-parameter-like onset at $T_\mathrm{N}$ and a phonon with a temperature-independent frequency to fit the temperature dependence of $L$ and $H$ quantitatively based on Eq.~(\ref{TDep}) (Methods). 
This fitting yields $T_\mathrm{N}=117$~K, bare magnon and phonon energies of $E_m=3.46$ THz and $E_p=3.49$ THz, and a coupling constant $g=0.062$ THz, consistent with previously reported values\autocite{Chen2024ThermalFePSe3,Ni2023SignaturesAntiferromagnets,Cui2023ChiralityAntiferromagnet}{}.
The temperature dependence thus confirms the coupled magnonic and phononic nature of $L$ and $H$. 

\begin{center}
\begin{figure}[H]
   \sbox0{\includegraphics[width=\textwidth]{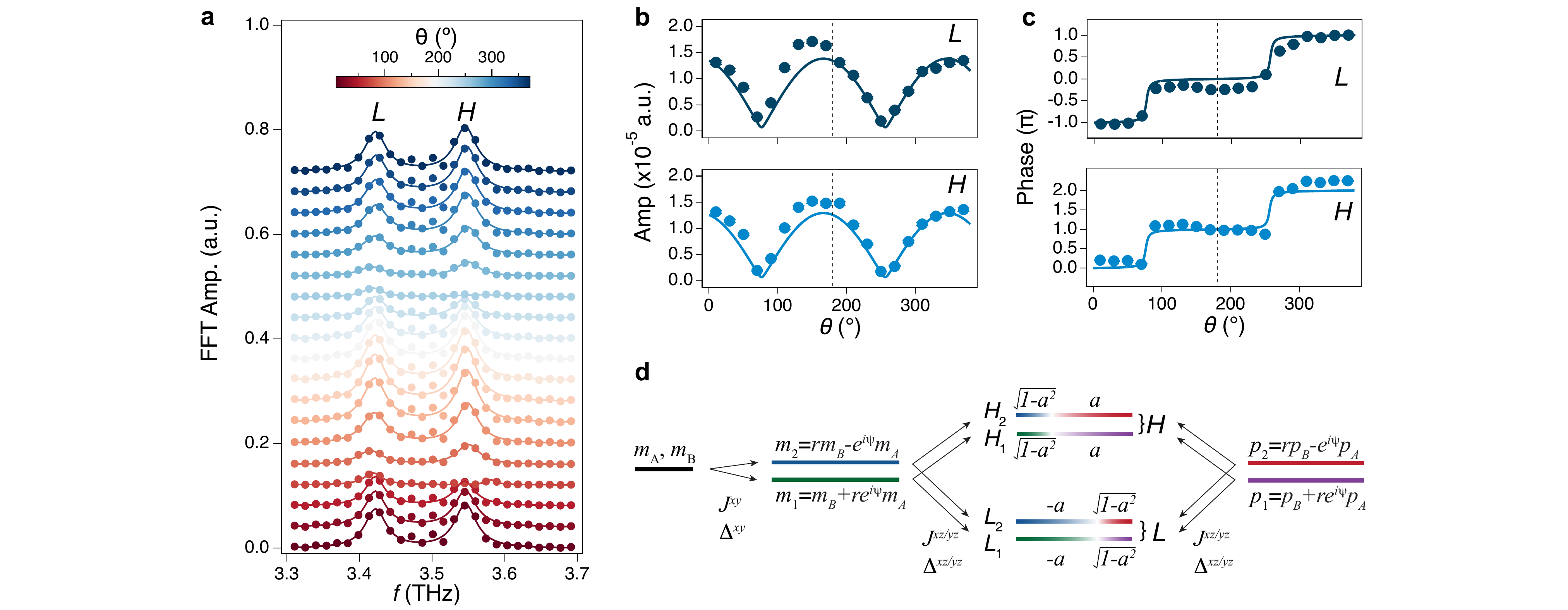}}
    \begin{minipage}{\wd0}
  \usebox0
  \captionsetup{labelfont={bf},labelformat={default},labelsep=period,name={Fig.}}
  \caption{\textbf{Phase-resolved polarimetry of the elliptical magnon polarons.} \textbf{a}, FFT spectra of $L$ and $H$ as a function of THz polarization angle $\theta$ obtained at the highest $E$ and 10 K. Curves are offset vertically for clarity. Fits with Lorentzians are overlaid with the data. \textbf{b,c}, $\theta$-dependencies of the amplitudes and phases of $L$ and $H$ obtained from the complex FFT. Solid lines are fits to Eq.~(\ref{AngleDepEq2}) and vertical dashed lines denote the position of $b$ axis. \textbf{d}, Schematic of the evolution of the energy levels in the microscopic model. Anisotropic exchange interactions $J^{xy}$,$J^{xz}$,$J^{yz}$ and non-Ising onsite anisotropies $\Delta^{xy}$,$\Delta^{xz}$,$\Delta^{yz}$ that induce splitting or mixture are denoted below the arrows. Composition of $m_1$, $m_2$, $p_1$, and $p_2$ in each state is colored green, blue, purple, and red, respectively.} 
  \label{Fig3}
\end{minipage}
\end{figure}  
\end{center}

\paragraph{}
Having established the coexistence of both magnonic and phononic components in $L$ and $H$ and confirmed the dominant linear magnetic dipolar excitation pathway, we now determine the ellipticity of $L$ and $H$ using phase-resolved polarimetry. 
We rotate the THz polarization angle $\theta$ from 0$^\circ$ to 360$^\circ$ while maintaining constant $E$, probe polarization and power, as well as sample orientation. 
$\theta$-dependent FFT amplitudes of $L$ and $H$, as shown in Fig.~\ref{Fig3}a, display synchronized evolution. 
To retrieve the coherent oscillation phase information, we conduct a complex FFT and fit the complex $L$ and $H$ peaks using two Lorentzians characterizing the damped MPs (Supplementary Note 7). 
A well-aligned twofold symmetry can be resolved in the fitted amplitudes and phases of both modes, with the maximal and minimal amplitudes deviating by $\sim20^\circ$ from the high-symmetry crystallographic axes (Figs.~\ref{Fig3}b,c).
This indicates the coexistence of both $A_g$ and $B_g$ components, reminiscent of the linear and elliptical MPs (Fig.~\ref{Fig1}). 
A close examination shows that the minimal amplitudes of both modes do not vanish, forming two nodeless ``waists". 
Furthermore, the phases of both MPs, though nearly constant within the two petals, smoothly change by $\pi$ within the angles where the waists appear. 
These signatures benchmark the elliptical nature of these MPs. 

\paragraph{}
To fit the polar patterns and retrieve the values of $r$ and $\psi$, we develop a quantitative model based on the microscopic Hamiltonian (see Methods). 
The leading-order magnetic Hamiltonian, including isotropic magnetic exchange interactions $J$ up to the third nearest neighbors and onsite Ising anisotropies $\Delta^z$, predict degenerate $B_g$ and $A_g$ magnons\autocite{Cui2023ChiralityAntiferromagnet,Luo2023EvidenceLimit}{}.
To fully capture the magnon-phonon coupling in $\mathrm{FePSe_3}$, we include subdominant anisotropic exchange interactions $J^{xy}$,$J^{xz}$,$J^{yz}$ and subleading onsite anisotropies $\Delta^{xy}$,$\Delta^{xz}$,$\Delta^{yz}$, which couple spin operators along different directions at different or same sites, respectively (e.g. $J^{xy}\hat{S}_i^x\hat{S}_j^y$ and $\Delta^{xz}\hat{S}_i^x\hat{S}_i^z$, where ${S}_i^z$ is the spin operator at site $i$ along the $z$-direction.).
Here, the $J^{xy}$ and $\Delta^{xy}$ terms modulate the magnon eigenvalues, inducing a weak splitting between the otherwise degenerate magnons and favoring a new elliptically polarized basis $m_1$ and $m_2$ with $r$ and $\psi$ values determined by the anisotropies (left panel of Fig.~\ref{Fig3}d). 
On the other hand, the $J^{xz/yz}$ and $\Delta^{xz/yz}$ terms determine the leading-order magnon-phonon coupling $g$, giving rise to the formation of MPs (middle panel of Fig.~\ref{Fig3}d)\autocite{Cui2023ChiralityAntiferromagnet}{}.
The selective magnon-phonon coupling ensures the identical composition of $p_1$ and $p_2$ as $m_1$ and $m_2$ with only two undetermined parameters $r$ and $\psi$.
Finally, $m_1$, $m_2$, $p_1$, and $p_2$ further mix into four nondegenerate states $L_1$, $L_2$, $H_1$, and $H_2$ (Fig.~\ref{Fig3}d), parametrized by a single parameter $a$ characterizing the composition of magnon and phonon in $L$ and $H$.
We then derive the expected $\theta$-dependencies of the magnetic dipole activity of $L$ and $H$ (see Methods for details) and globally fit their amplitudes and phases based on Eq.~(\ref{AngleDepEq2}).
The fitting and data exhibit a remarkable agreement with $\psi=0.93\pi$, $r=1.66$, and $a=0.73$, highlighting the elliptical nature of the MPs (Fig.~\ref{Fig3}b).



\begin{center}
\begin{figure}[H]
   \sbox0{\includegraphics[width=\textwidth/2]{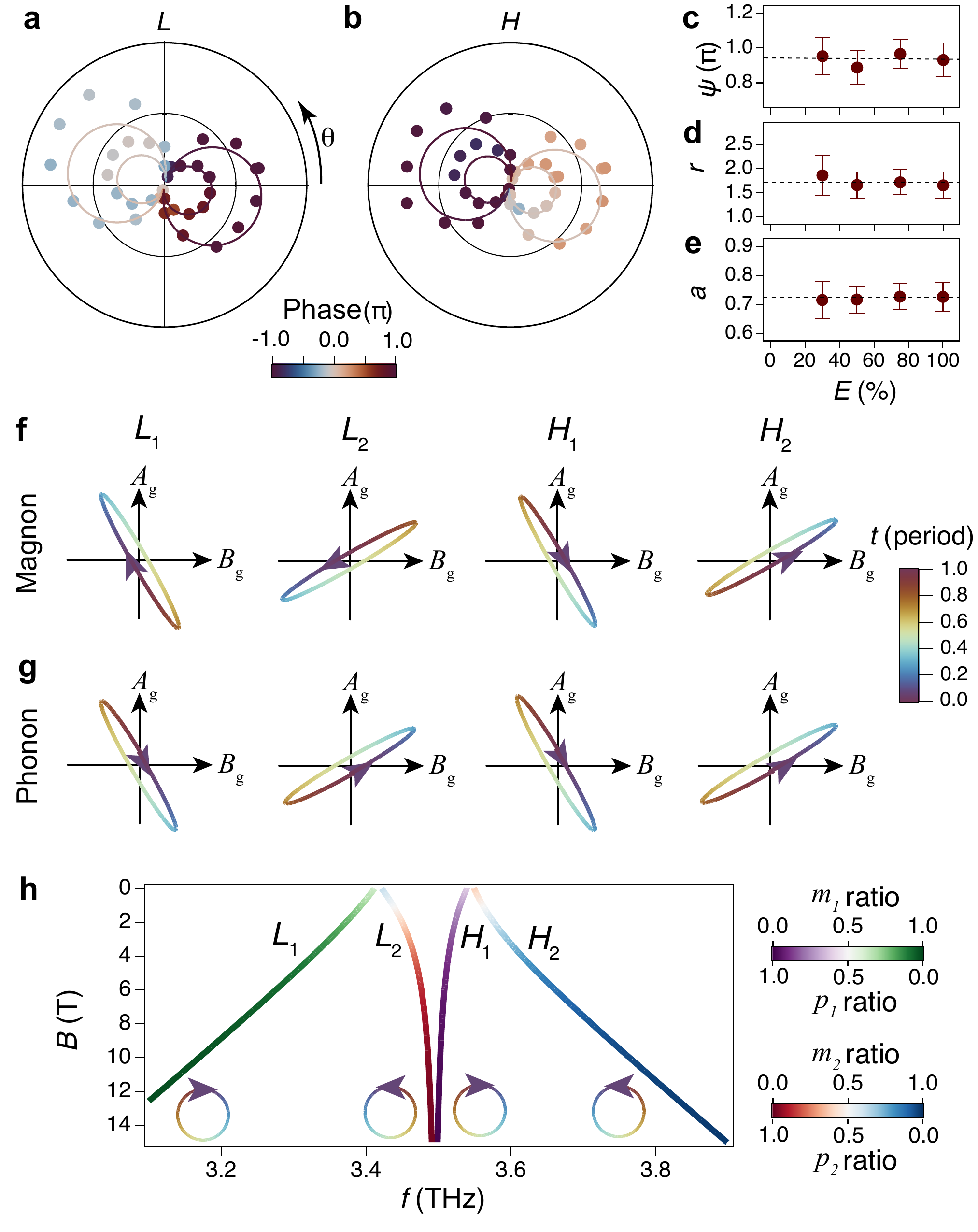}}
    \begin{minipage}{\wd0}
  \usebox0
  \captionsetup{labelfont={bf},labelformat={default},labelsep=period,name={Fig.}}
  \caption{\textbf{$E$-dependent polarimetry and reconstructed trajectories of the magnon-polarons.} \textbf{a,b}, Phase-resolved polarimetry of $L$ and $H$ measured at $50\%$ and $100\%$ of the highest $E$ and 10 K. \textbf{c-e}, $E$-dependencies of $\psi$, $r$, and $a$. Dashed lines denote their $E$-averaged values. \textbf{f,g}, Reconstructed trajectories of the net magnetization (magnon) and atomic displacement (phonon) of the four elliptical MPs $L_1$, $L_2$, $H_1$, and $H_2$ in the $\{B_g,A_g\}$ coordinates. Time $t$ is color coded periodically. \textbf{h}, Calculated magnetic field dependence of the energies of the four elliptical MPs. Magnon and phonon compositions are color coded. In the high-field limit, each branch features purely circular motions with $r=1,\psi=\frac{\pi}{2}$.} 
  \label{Fig4}
\end{minipage}
\end{figure}  
\end{center}

\paragraph{}
By repeating these polarimetry measurements of $L$ and $H$ at different $E$ values and fitting the phase-coded amplitude polar patterns (Figs.~\ref{Fig4}a,b), 
we find a clear independence in $E$ of the fitted values of $\psi$, $r$, and $a$ (Figs.~\ref{Fig4}c-e). 
This observation agrees with the facts that the linear magnetic dipolar channel is consistently the dominant excitation mechanism for MPs and that $\psi$, $r$, and $a$ reflect the intrinsic properties of the elliptical MPs. 
Using the averaged values of $\psi=0.93\pi$, $r=1.72$, and $a=0.72$, we can reconstruct the magnonic and phononic trajectories of all four modes $L_1$, $L_2$, $H_1$, and $H_2$, where prominent elliptically polarized rotations with opposite helicities can be resolved (Figs.~\ref{Fig4}f,g).

\paragraph{}
We now compare these experimentally obtained values to theoretical predictions by performing first-principles calculations to compute the exchange interactions, $J$'s, anisotropy $\Delta^z$, magnon and phonon frequencies, and electron-phonon coupling constants (Supplementary Note 8). 
Based on these parameters, our calculations yield a value of $a^2=0.57$, in reasonable agreement with the values obtained from the $T$-dependent measurements ($a^2=0.61$) and the polarimetry measurements ($a^2=0.53$).
The values of $r$ and $\psi$ depend sensitively on $J^{xy}$ and $\Delta^{xy}$, and accurate determination of these subdominant terms based on first-principles calculations is highly challenging. 
Based on an order-of-magnitude estimate, our theory predicts the value of $r$ close to 1 and $\psi$ several degrees deviated from $\pi$, both of which are qualitatively consistent with our experimentally acquired values.  
Accordingly, the reasonable agreement between theory and experiment validates our microscopic interpretation.

\paragraph{}
In conclusion, our results demonstrate the existence of nondegenerate elliptical MPs at the center of the Brillouin zone and the feasibility of exciting them with a resonant linearly polarized THz light. 
We note that our findings are different from a recent theoretical proposal to generate chiral phonons using linearly polarized light in paramagnets, where a slow build-up time of net polarization and a threshold electric field are required\autocite{Ren2024Light-DrivenParamagnets}{}. 
Our work also contrasts with previous experimental achievements, where one or more linearly polarized pulses driving two orthogonal phonons are used to induce ``artificial" ellipticity that is highly sensitive to the polarization of the pump or the delay between the pumps \autocite{Nova2017AnPhonons,Wefers1998OpticalQuartz,Katsuki2013All-opticalBismuth}{}.
Our work further provides a unique polarimetry-based method to reconstruct the elliptical phonon trajectories, expanding the current scope of the burgeoning field \autocite{Tauchert2022PolarizedDemagnetization,Ueda2023ChiralX-rays}.
We anticipate that this methodology may be applied to other systems with strong phonon-magnon coupling and anisotropic magnetic interactions.
Our theory also predicts that the application of a large enough out-of-plane magnetic field $B$ can not only significantly decouple the magnons and phonons, but also make both magnons and phonons perfectly circularly polarized, i.e. $r=1$, $\psi=\frac{\pi}{2}$ (Figs.~\ref{Fig4}h). 
Further THz polarimetry measurements complemented by a tunable external magnetic field shoule be able to reveal such an elliptical-to-circular MP transition.
With the development of state-of-the-art narrowband THz sources with accurately tunable phases, selective and enhanced excitation of one elliptically polarized MP to generate a large effective magnetic moment is also within reach\autocite{Juraschek2022GiantParamagnets,Luo2023LargeHalides}.

\section*{Methods}

\paragraph{Sample preparation\\} 
$\mathrm{FePSe_3}$ single crystals were synthesised using the chemical vapour transport method. The starting materials iron (99.999\%), phosphorus (99.999\%) and selenium (99.999\%) powders were mixed in a stoichiometric ratio, with iodine added as a transport agent. These materials were sealed in a quartz tube, evacuated to $3\times10^{-2}$ torr, and placed in a two-zone furnace (Ajeon Instrument, Korea). The synthesis temperature was set to 750 $^\circ$C/720 $^\circ$C for one week, followed by 500 $^\circ$C/450 $^\circ$C for 2 days, and then 150 $^\circ$C/120 $^\circ$C for 6 hours in the hot/cold zone, respectively. Finally, the materials were cooled down to room temperature. The resulting single crystals were characterized using energy dispersive X-ray spectrometry and powder X-ray diffraction. Single crystals were cleaved along the [001] direction immediately before the experiment and was then kept in a vacuum with a pressure lower than $10^{-6}$ torr at low temperatures.

\paragraph{THz experiment\\}
A detailed description can be found in Supplementary Note 2. The broadband THz pump is generated by pumping a N-benzyl-2-methyl-4-nitroaniline crystal with the 1300 nm optical parametric amplifier output seeded by a Ti:Sapphire amplifier at a repetition rate of 1 kHz. A much weaker 800 nm pulse from the amplifier is focused on the sample to probe the THz-induced polarization ellipticity change with the balanced detection scheme using a pair of photodiodes. The THz and 800 nm pulses are both normally incident on the (001) face of the sample.

\paragraph{Polarimetry fitting\\} 
We can quantitatively rationalize the selective coupling between $m_1$, $m_2$, $p_1$, and $p_2$ in a matrix form:
\begin{equation}\label{T_{1212}}
    H=\begin{pmatrix}
           E_{m1} & 0 & g_1 & 0  \\
           0 & E_{m2} & 0 & g_2  \\
           g_1 & 0 & E_{p1} & 0  \\
           0 & g_2 & 0 & E_{p2}  \\
    \end{pmatrix},
\end{equation}
The energies of the four eigenstates and their corresponding eigenstates in $\{m_1,m_2,p_1,p_2\}$ basis can be expressed as:
\begin{equation}\label{EigenValuesandStates}
\begin{split}
&E_{L_1}=\frac{1}{2}(E_{m_1} + E_{p_1} - \sqrt{4g_1^2+(E_{m_1} - E_{p_1})^2}),\hspace{0.25in} \Psi_{L_1}=\{-a_1, 0, \sqrt{1-a_1^2}, 0\},\\
&E_{L_2}=\frac{1}{2}(E_{m_2} + E_{p_2} - \sqrt{4g_2^2+(E_{m_2} - E_{p_2})^2}),\hspace{0.25in} \Psi_{L_2}=\{0, -a_2, 0, \sqrt{1-a_2^2}\},\\
&E_{H_1}=\frac{1}{2}(E_{m_1} + E_{p_1} + \sqrt{4g_1^2+(E_{m_1} - E_{p_1})^2}),\hspace{0.25in} \Psi_{H_1}=\{\sqrt{1-a_1^2}, 0, a_1, 0\},\\
&E_{H_2}=\frac{1}{2}(E_{m_2} + E_{p_2} + \sqrt{4g_2^2+(E_{m_2} - E_{p_2})^2}),\hspace{0.25in} \Psi_{H_2}=\{0, \sqrt{1-a_2^2}, 0, a_2\},\\
\end{split}
\end{equation}
where $E_{m_1/m_2/p_1/p_2}$ are the uncoupled energies of different modes and $g_{1/2}$ are the selective coupling constants between $m_{1/2}$ and $p_{1/2}$, respectively.
Note that the form of the eigenstates underscores the selective coupling between $m_1$ ($m_2$) and $p_1$ ($p_2$).
Since the difference between $E_m$'s, $E_p$'s, and $g$'s are much smaller than their averages, which are confirmed by both our experimental resolution and previous Raman scattering measurements \autocite{Cui2023ChiralityAntiferromagnet,Jana2023In-planeScattering,Luo2023EvidenceLimit}{}, we can assume $E_{m_1}\sim E_{m_2}=E_m$, $E_{p_1}\sim E_{p_2}=E_p$, and $g_1\sim g_2=g$ and so $E_{L_1}\sim E_{L_2}$, $E_{H_1}\sim E_{H_2}$, and $a_1\sim a_2=a$. 
Indeed, our experiment cannot distinguish $L_1$ ($H_1$) from $L_2$ ($H_2$) but only resolve their combination $L$ ($H$) mode. 
With these assumptions, the eigenfunctions $\Psi$'s, which represent the composition of $m_1$, $m_2$, $p_1$, and $p_2$ in $L_1$, $L_2$, $H_1$, and $H_2$, can be parametrized by a single parameter $a=\frac{\sqrt{4 g^2+(E_m-E_p)^2}-(E_m-E_p)}{\sqrt{4 g^2+(\sqrt{4 g^2+(E_m-E_p)^2}-(E_m-E_p))^2}}$. 
Substituting $m_1=m_B+r e^{i\psi}m_A$, $m_2=rm_B- e^{i\psi}m_A$, $p_1=p_B+r e^{i\psi}p_A$, and $p_2=rp_B- e^{i\psi}p_A$, we can thus express $L$ and $H$ as:
\begin{equation}\label{ComponentEq}
\begin{split}
\Psi_{L}&=-a(1+r)m_{B}-a(r-1)e^{i\psi}m_{A}+\sqrt{1-a^2}(1+r)p_{B}+\sqrt{1-a^2}(r-1)e^{i\psi}p_{A},\\
\Psi_{H}&=\sqrt{1-a^2}(1+r)m_{B}+\sqrt{1-a^2}(r-1)e^{i\psi}m_{A}+a(1+r)p_{B}+a(r-1)e^{i\psi}p_{A}.
\end{split}
\end{equation}
Since the linear excitation channel dominates, our measurements reflect the components of $m_B$ and $m_A$, which show $\cos\theta$ and $\sin\theta$ dependencies, respectively (Supplementary Note 7). Therefore, the predicted $\theta$-dependencies of the measured polarimetry are:
\begin{equation}\label{AngleDepEq2}
\begin{split}
A_{L}(\theta) &\propto -a(1+r)\cos{\theta}-a(r-1)e^{i\psi}\sin{\theta}, \\
A_{H}(\theta) &\propto \sqrt{1-a^2}(1+r)\cos{\theta}+\sqrt{1-a^2}(r-1)e^{i\psi}\sin{\theta},
\end{split}
\end{equation}
These expressions have two implications. 
First, the polarimetry of both the amplitudes and phases of $L$ and $H$ is determined by three parameters, $\psi$, $r$, and $a$. 
Second, we expect the same $\theta$-dependence of $L$ and $H$ with a scaling factor reflecting the value of $a$. 

\paragraph{Temperature dependence fitting\\} 
Assuming that $E_{m_1}=E_{m_2}=E_m$, $E_{p_1}=E_{p_2}=E_p$, and $g_1=g_2=g$, $E_{L_1}=E_{L_2}=E_L$ and $E_{H_1}=E_{H_2}=E_H$ can be expressed as:
\begin{equation}\label{TDep}
\begin{split}
&E_{L}(T)=\frac{1}{2}(E_{m}(T) + E_{p}(T) - \sqrt{4g^2+(E_{m}(T) - E_{p}(T))^2}),\\
&E_{H}(T)=\frac{1}{2}(E_{m}(T) + E_{p}(T) + \sqrt{4g^2+(E_{m}(T) - E_{p}(T))^2}).
\end{split}
\end{equation}
By further assuming a phenomenological temperature dependence of $E_m(T)=E_m\tanh(a\sqrt{T_\mathrm{N}/T - 1})$ and temperature independent $E_p(T)=E_p$ and $g$, we can fit the temperature dependence of $E_{L}(T)$ and $E_{H}(T)$ and obtain the values of $E_m$, $E_p$, $g$, and $T_\mathrm{N}$, which are discussed in the main text.

\section*{Data Availability}
\noindent The datasets generated and/or analysed during the current study are available from the corresponding author on request.

\section*{Code Availability}
\noindent The code used for the current study are available from the corresponding author on request.

\printbibliography

\section*{Acknowledgments}
\noindent We thank Jacob Pettine, Alexander von Hoegen and Zhuquan Zhang for helpful discussions. We acknowledge the support from the US Department of Energy, Materials Science and  Engineering  Division,  Office  of  Basic  Energy  Sciences  (BES  DMSE)  (data taking and analysis), Gordon and Betty Moore Foundation’s EPiQS Initiative grant GBMF9459 (instrumentation and manuscript writing), and the MIT-Israel Zuckerman STEM Fund. E.V.B. acknowledges funding from the European Union's Horizon Europe research and innovation programme under the Marie Skłodowska-Curie grant agreement No 101106809. A.R. was supported by the Cluster of Excellence Advanced Imaging of Matter (AIM), Grupos Consolidados (IT1249-19), SFB925, “Light Induced Dynamics and Control of Correlated Quantum Systems,” and the Max Planck Institute New York City Center for Non-Equilibrium Quantum Phenomena. D.M.J. is supported by Tel Aviv University. The work at Seoul National University was supported by the Leading Researcher Program of Korea’s National Research Foundation (Grant No. 2020R1A3B2079375).

\section*{Author Contributions}
\noindent H.N., T.L., and B.I. performed the measurements, analyzed the data, and interpreted the results. H.N. developed the theory with help from T.L. and E.V.B. J.P. and J.K. synthesized and characterized the single crystals under the supervision of J.-G.P. H.N. wrote the manuscript with critical input from T.L., B.I., E.V.B., D.M.J., A.R., N.G., and all other authors. The project was supervised by N.G.

\section*{Competing Interests}
\noindent The authors declare no competing interests.

\newpage

\newpage

\newpage

\newpage

\mbox{~}
\clearpage
\end{document}